\documentclass[aps,prl,twocolumn,nobibnotes,superscriptaddress]{revtex4-2}

\usepackage{xcolor}
\usepackage{hyperref}
\hypersetup{
    colorlinks=true,
	linkcolor={cyan!50!black},
	citecolor={red!50!black},
	urlcolor={red!40!black}
}

\usepackage{graphicx}
\renewcommand{\figurename}{Fig.}

\usepackage{amssymb}
\usepackage{amsmath}
\usepackage{bm}

\newcommand{\para}{||}
\newcommand{\x}{{\sf X}}
\newcommand{\Bx}{B_\x}
\newcommand{\amp}{\chi}
\newcommand{\displ}{\chi}
\newcommand{\Bpara}{B_{\para}}
\newcommand{\Bperp}{B_{\perp}}

\newcommand{\gpara}{g_{\para}}
\newcommand{\gperp}{g_{\perp}}
\newcommand{\Pflip}{P_{\Downarrow}} 
\newcommand{\muB}{\mu_\mathrm{B}}
\newcommand{\rest}{_\mathrm{rest}}

\newcommand{\fspin}{f_{\rm spin}}
\newcommand{\fosci}{f_{\rm osci}}
\newcommand{\Pspin}{P_{\rm spin}}
\newcommand{\Posci}{P_{\rm osci}}

\newcommand{\EDSR}{{\rm EDSR}}
\newcommand{\fEDSR}{f_{\rm EDSR}}
\newcommand{\fm}{f_{\rm m}}
\newcommand{\sigth}{\sigma_3}
\newcommand{\ind}[1]{_{#1}}

\newcommand{\affeng}{Dept. of Engineering Science, University of Oxford, Oxford OX1 3PJ, United Kingdom.}
\newcommand{\affexe}{Dept. of Physics and Astronomy, University of Exeter, Stocker Road, EX4 4QL, United Kingdom.}
\newcommand{\affgren}{Univ. Grenoble Alpes, CNRS, Grenoble INP, Institut N\'eel, 38000 Grenoble, France.}
\newcommand{\affmat}{Dept. of Materials, University of Oxford, Oxford OX1 3PH, United Kingdom.}
\newcommand{\affchal}{Dept. of Microtechnology and Nanoscience (MC2), Chalmers University of Technology, S-412 96 G\"oteborg, Sweden.}
\newcommand{\affmad}{Dept. of Structure of Matter, Thermal Physics and Electrodynamics and GISC, Universidad Complutense de Madrid, Pl. de las Ciencias 1. 28040 Madrid, Spain.}
\newcommand{\affpots}{Institute of Physics and Astronomy, University of Potsdam, 14476 Potsdam, Germany.}
\newcommand{\affmaju}{MajuLab, National University of Singapore, 117543 Singapore.}
\newcommand{\affbuda}{Dept. of Theoretical Physics, Budapest University of Technology and Economics, M\H{u}egyetem rkp. 3., H-1111 Budapest, Hungary.}
\newcommand{\affhun}{HUN-REN--BME Quantum Dynamics and Correlations Research Group, M\H{u}egyetem rkp. 3., H-1111 Budapest, Hungary}

\begin{document}

\title{Coupling a single spin to high-frequency motion}

\author{Federico Fedele}
\thanks{These authors contributed equally to this work.}
\affiliation{\affeng}
\author{Federico Cerisola}
\thanks{These authors contributed equally to this work.}
\affiliation{\affexe}
\affiliation{\affeng}
\author{Lea Bresque}
\affiliation{\affgren}
\author{Florian Vigneau}
\affiliation{\affmat}
\author{Juliette Monsel}
\affiliation{\affchal}
\author{Jorge Tabanera}
\affiliation{\affmad}
\author{Kushagra Aggarwal}
\affiliation{\affmat}
\author{Jonathan Dexter}
\affiliation{\affmat}
\author{Sofia Sevitz}
\affiliation{\affpots}
\author{Joe Dunlop}
\affiliation{\affexe}
\author{Alexia Auffeves}
\affiliation{\affgren}
\affiliation{\affmaju}
\author{Juan Parrondo}
\affiliation{\affmad}
\author{Andras Palyi}
\affiliation{\affbuda}
\affiliation{\affhun}
\author{Janet Anders}
\affiliation{\affexe}
\affiliation{\affpots}
\author{Natalia Ares}
\email{natalia.ares@eng.ox.ac.uk}
\affiliation{\affeng}

% Contact:\\ \href{mailto:federico.fedele@eng.ox.ac.uk}{federico.fedele@eng.ox.ac.uk} and
% \href{mailto:f.cerisola@exeter.ac.uk}{f.cerisola@exeter.ac.uk}

\begin{abstract}
Coupling a single spin to high-frequency mechanical motion is a fundamental bottleneck of applications such as quantum sensing~\cite{Chui2004,Wrachtrup2008,Kolkowitz2012,Degen2022}, intermediate and long-distance spin-spin coupling~\cite{Lukin2010}, and classical and quantum information processing~\cite{Lukin2021}. 
%Previous experiments have observed such coupling in low-frequency mechanical resonators, which are mostly confined to the classical regime, such as diamond cantilevers~\cite{Kolkowitz2012,Pigeau2015,Barfuss2015,Jayich2017,Ovartchaiyapong2014}.
%{\color{blue}Previous experiments have observed such coupling in low-frequency mechanical resonators such as diamond cantilivers~\cite{Kolkowitz2012,Pigeau2015,Barfuss2015,Jayich2017,Ovartchaiyapong2014}. However, these systems are mostly confined to the classical regime and thus cannot be applied to coherent state transfer between the spin and the resonator, quantum control of the mechanical resonator, and quantum limited measurements.}
Previous experiments have only shown single spin coupling to low-frequency mechanical resonators, such as diamond cantilevers  ~\cite{Kolkowitz2012,Pigeau2015,Barfuss2015,Jayich2017,Ovartchaiyapong2014}. 
High-frequency mechanical resonators, having the ability to access the quantum regime, open a range of possibilities when coupled to single spins, including readout and storage of quantum states.
%coherent state transfer between the spin and the mechanics and quantum control of the mechanical resonator
Here we report the first experimental demonstration of spin-mechanical coupling to a high-frequency resonator. 
We achieve this all-electrically on a fully suspended carbon nanotube device.
A new mechanism gives rise to this coupling, which stems from spin-orbit coupling, and it is not mediated by strain~\cite{Payli2012}.
We observe both resonant and off-resonant coupling as a shift and broadening of the electric dipole spin resonance (EDSR), respectively.
We develop a complete theoretical model taking into account the tensor form of the coupling and non-linearity in the motion. 
Our results propel spin-mechanical platforms to an uncharted regime. 
The interaction we reveal provides the full toolbox for promising applications ranging from the demonstration of macroscopic superpositions~\cite{Yiwen2023}, to the operation of fully quantum engines~\cite{Llobet2015,Culhane2022,Josefsson2018}, to quantum simulators~\cite{Tacchino2018}.
\end{abstract}

% \vspace{5ptw}
%\section*{Main text}

\maketitle

The ability to couple a single spin to mechanical motion emerges as a requirement for quantum technological applications~\cite{Chui2004,Wrachtrup2008,Lukin2010,Kolkowitz2012,Awschalom2019,Lukin2021,Degen2022}. Mechanical resonators can extend over macroscopic distances and exhibit large quality factors. This means that the coupling between spin and motion can be used to connect well-separated locations on chip and to store quantum information, enabling hybrid quantum networks~\cite{Polzik2021}. 
Unlike most of the alternatives for intermediate and long-range spin-spin coupling, mechanical resonators can exhibit longitudinal parametric coupling to spins~\cite{Pigeau2015}, enabling low power entangling operations~\cite{Yacobi2022} and fast quantum-non-demolition readout of the spin states~\cite{Blais2015}. Spin mechanical coupling is highly desired to enable teleportation and entanglement swapping of macroscopic states of motion~\cite{Treutlein2020}.
%Unlike most of the alternatives for long-range spin-spin coupling and quantum memories based on superconducting materials, mechanical resonators are also unaffected by large magnetic fields. 

Recent advances in magneto-mechanical sensing are also finding practical applications in biology and biomedicine~\cite{Park2023,Rahmer2023}.
%From a fundamental perspective, the coupling between spins and mechanics is highly desired to enable teleportation and entanglement swapping of macroscopic states of motion~\cite{Pigeau2015,Treutlein2020}.{\color{blue}From a fundamental perspective, the coupling between spins and mechanics closely mirrors the principles of quantum electrodynamics (QED) and it opens the opportunity for key experiments that explore the foundations of quantum mechanics~\cite{Pigeau2015}.}
From a fundamental perspective, the coupling between spins and mechanics is key to the exploration of the quantum-to-classical transition.
These prospects have motivated extensive theoretical and experimental work to achieve coupling between single spins and mechanical resonators that can be operated in the quantum regime~\cite{Lukin2009,Jayich2017,Morello2021}.

%%%%%%%%%%%%%%%%%%%%%
\begin{figure*}[t!]
\includegraphics[trim=0.2cm 0.2cm 0.2cm 0.0cm,clip,width=0.98\textwidth]{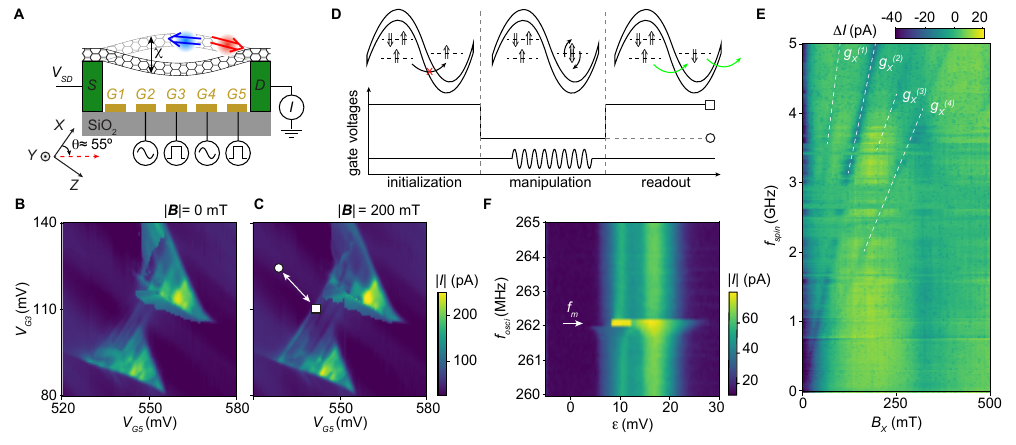}
\caption[]{
\textbf{Determination of spin-valley and mechanical frequencies.}
(\textbf{A}) 
The carbon nanotube is suspended between source and drain contacts, and over five gate electrodes, G1-G5. 
We measure a current $I$ through the device against a source-drain voltage $V_{SD}$. 
A magnetic field $\Bx$ is applied along the $\x$ direction, and the CNT is oriented at an angle $\theta$ of approximately $55^\circ$ against the $\x$ axis. A double dot confined along the CNT (see red and blue double-arrows) allows us to define a spin-valley qubit and to use Pauli spin blockade for readout. The CNT motion oscillates with an amplitude $\amp$ as a response to a drive tone at frequency $\fosci$ in one of the gates.
(\textbf{B}) and (\textbf{C}) show the absolute value of the current $|I|$, measured as a function of gate voltages $V_{G3}$ and $V_{G5}$ at a Pauli-blocked transition. Other voltages are $V_{SD} = -4$~mV, $V_{G1}=V_{G2}=V_{G4}=0$~V, and the applied field magnitude is $|\bm{B}|=0$~mT and $|\bm{B}|=200$~mT, respectively. The square (circle) indicates the initialization/readout (manipulation) gate voltages and the arrow marks the pulse axis, see schematic \textbf{D}. 
(\textbf{D}) Pulse scheme for qubit manipulation. Upper panel: Double-dot occupations indicating that Pauli spin blockade makes the current sensitive to the spin-valley state (double-arrows). Lower panel: Cycle of gate voltage pulses for qubit initialisation, qubit manipulation, and detection of EDSR. The qubit is driven with a microwave pulse at frequency $\fspin$.
(\textbf{E}) Resonant signal $\Delta I$ as a function of drive frequency $\fspin$ and applied field $\Bx$. The pulse cycle used is that in panel \textbf{D} and the microwave power $\Pspin$ was $-33$~dBm at the device (here applied to G4). The average current in each row is subtracted to highlight the EDSR resonances ($\fEDSR$). Resonances appear as dark lines (dashed white lines are a guide to the eye). Land\'e $g$-factor values corresponding to these resonances are shown.
(\textbf{F}) Observation of the nanotube's mechanical frequency ($\fm$), identified at ca. $261.9$~MHz, by measuring the current $I$ as a function of double dot detuning ($\varepsilon$) while sweeping $\fosci$. The drive power $\Posci$ (applied to G4) is $-33$~dBm  at the device and the field is $\Bx=200$~mT. Note that the detuning start point and axis are different than those indicated in panel \textbf{C}. No EDSR excitation was applied for this measurement.} 
\label{fig_1}
\end{figure*}
%%%%%%%%%%%%%%%%%%%%%

The interaction between single spins and mechanics has been demonstrated for a single NV spin-qubit probed by a magnetized AFM cantilever~\cite{Kolkowitz2012,Hong2012}, as well as in monolithic platforms, where single NV spin-qubits are embedded in cantilevers and semiconductor nanowires~\cite{Ovartchaiyapong2014,Barfuss2015,Teisser2014,Arcizet2011,Pigeau2015}, and with an InGaAs quantum dot~\cite{Carter2018} hosted in a GaAs cantilever. 
These mechanical oscillators have resonant frequencies below ~6 MHz and thus high phonon occupancies even at cryogenic temperatures. 
Cooling mechanical motion can be achieved using a variety of techniques~\cite{Aspelmeyer2014}, including laser cooling~\cite{Delic2020}. 
However, these techniques are often incompatible with qubit operation, for example due to substrate heating~\cite{Millen2014}. 
The challenge resides in realising spin-to-motion coupling in a mechanical resonator exhibiting large quality factors and a large resonance frequency, so that the quantum regime can be accessed at dilution refrigerator temperatures.
Due to their small mass and unique material properties, carbon nanotubes (CNTs) can realise resonators with quality factors of over a million \cite{Moser2014} and resonant frequencies of up to 39~GHz \cite{Laird2012}. This type of device allows for quantum dots to be defined electrostatically~\cite{Ilani2013,Ilani2014}. The coupling between single-electron tunneling and the CNT motion was recently estimated and it was found to reach the ultrastrong coupling regime~\cite{Vigneau2022,Samanta2023}. Strong coupling of single spins to mechanical motion was predicted~\cite{Payli2012}, but never observed.

Here, we report on the first observation of coupling between a single spin and mechanical motion at radio frequencies. The single spin is confined in a quantum dot which is electrostatically defined in a fully suspended carbon nanotube. It is driven by EDSR and interacts with the nanotube's motion via spin-orbit coupling. We observe this coupling when the Larmor and mechanical frequencies are both, resonant and off-resonant. We develop a theoretical
model that captures the dependence of the gyromagnetic tensor on the carbon nanotube displacement, including non-linearities in the nanotube's motion. Our model can accurately reproduce our observations. The demonstration of spin-mechanical coupling unveils a range of new opportunities, spanning both fundamental and practical applications of quantum mechanics.

%%%%%%%%%%%%%%%%%%%%%
\begin{figure*}[t]
\includegraphics[trim=0.2cm 0.1cm 0.1cm 0.2cm,clip,width=1\textwidth]{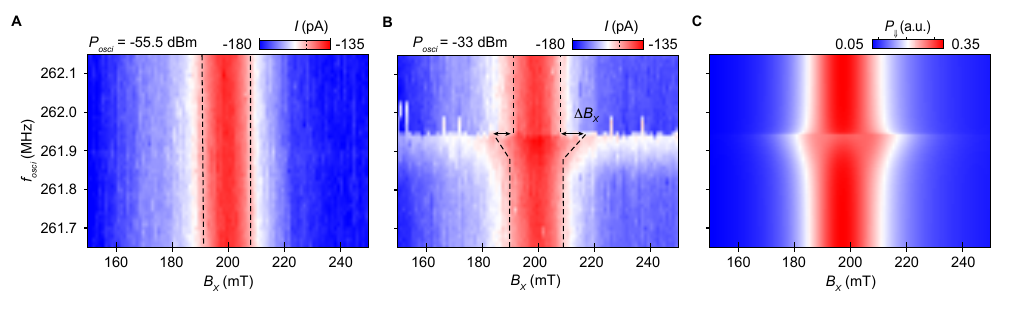}
\caption[]{
\textbf{Off-resonant coupling between a spin-valley qubit and mechanical motion.} 
(\textbf{A-B}) 
Measured EDSR resonance at a field $\Bx\approx 200$~mT corresponding to a drive at $\fspin \sim 5.1$~GHz, while the carbon nanotube motion is excited (at G2) at frequencies $\fosci$ in the range [261.7, 262.1]~MHz. The resonance is the one with $g$-factor $g^{(2)}_\x$ in Fig.~\ref{fig_1}. Here the spin manipulation is performed using a continuous microwave drive (at G4) with power $\Pspin=-34.5$~dBm. 
Panels \textbf{A} and \textbf{B} correspond to different mechanical drive powers $\Posci$, -55.5 dBm and -33 dBm, respectively. For the larger value of $\Posci$ (panel \textbf{B}), we observe EDSR resonance broadening ($\Delta \Bx^\EDSR$ indicated by arrows) at $\fm \sim261.9$~MHz, the frequency corresponding to the mechanical resonance. To quantify this broadening, we pinpoint the values of $\Bx$ corresponding to a reduction of $1/\sqrt{2}$ of EDSR current value (dashed black lines). Panel \textbf{C} shows the calculated transition probability $P_{\Downarrow}$ from the blocked to the unblocked state, time-averaged until the steady state is reached.  The transition probability is obtained by solving the dynamics of Hamiltonian \eqref{eq:Hexpansion} up to first order in $\chi$ (with $\lambda\ind{1} = 433$ neV/nm) plus the EDSR driving tone, and including decoherence effects due to the environment (see Methods for details). Our simulation reproduces the EDSR broadening observed in panel \textbf{B}. 
}
\label{fig_2}
\end{figure*}
%%%%%%%%%%%%%%%%%%%%%

The CNT is stamped across metallic contact electrodes to give a vibrating segment of length $\sim 900$~nm, and is measured at a temperature of $45$~mK. Voltages applied to five finger gates beneath the nanotube (labeled $V_{G1}$–$V_{G5}$) are used to form a double quantum dot (DQD) electrostatically (Fig.~\ref{fig_1}\textbf{A}). 
A voltage bias $V_{SD}$ is applied between the leads to drive a DC current $I$.
Both p-p and p-n double-dot configurations are accessible. We focus on the charge transition (N,1) $\to$ (N+1,0) in a p-n configuration. Since a  p-n bandgap only seems present for the right dot (see Methods), we can only assign absolute charge occupations to the right dot and relative charge occupations $N$ to the left dot.

With an external magnetic field $\Bx$, Pauli spin blockade is observed as an enhancement of the triangle baseline due to selection rules on spin and valley states (Fig.~\ref{fig_1}\textbf{B},\textbf{C}). Pauli spin blockade is identified by a suppression instead of an enhancement of the triangle baseline, with the latter signature found only in systems with strong-spin orbit coupling~\cite{Pfund2007,Danon2009,Nadj-Perge2010,Li2015}.
A cycle of gate voltage pulses applied to G3 and G5, see Fig.~\ref{fig_1}{\bfseries D}, is used to define and control a spin-valley qubit using electrically driven spin resonance (EDSR) \cite{Laird2013}. First, an effective triplet state is initialized by configuring the double dot in Pauli spin blockade. Then, G3 and G5 are pulsed to set the double dot in Coulomb blockade, and a microwave burst applied to either G2 or G4 manipulates the spin-valley state. Gate voltages G3 and G5 are pulsed back to the Pauli spin blockade configuration. If the spin-valley state was flipped during the microwave burst, Pauli spin blockade is lifted and the current changes. 
The pulse cycle is set at 1 $\mu$s. The resulting change in current $\Delta I$ as a function of the frequency of the microwave burst $\fspin$ and magnetic field $\Bx$ is shown in Fig.~\ref{fig_1}\textbf{E}. Dips in $\Delta I$ appear as diagonal lines indicating a resonance when $\fspin$ matches the qubit frequency $\fEDSR$. 
From the slopes of these diagonal lines, we extract effective $g$-factors: $g^{(1)}_\x =$ ($3.30\pm 0.30$), $g^{(2)}_\x =$ ($1.84\pm 0.02$), $g^{(3)}_\x =$ ($1.31\pm 0.06$) and $g^{(4)}_\x =$ ($0.94\pm 0.01$). 
Errors reflect the uncertainty of a linear fit to the EDSR resonances.
These multiple EDSR resonances $g^{(j)}_\x$ can be attributed to transitions within different subsets of spin-valley states ($g^{(1)}_\x$ and $g^{(3)}_\x$), as well as higher harmonics ($g^{(2)}_\x$ and $g^{(4)}_\x$).
We actuate vibrations by injecting a radio-frequency (RF) tone with driving power $\Posci$ to either G2 or G4. An abrupt change in $I$ is observed when the frequency of the mechanical drive tone $\fosci$ is close to the resonance frequency of the mechanical resonator at ca. $f_m=261.9$~MHz (Fig.~\ref{fig_1}\textbf{F}).

%%%%%%%%%%%%%%%%%%%%%
\begin{figure*}[t]
\flushleft \includegraphics[trim=0.1cm 0.1cm 0.1cm 0.1cm,clip,width=1\textwidth]{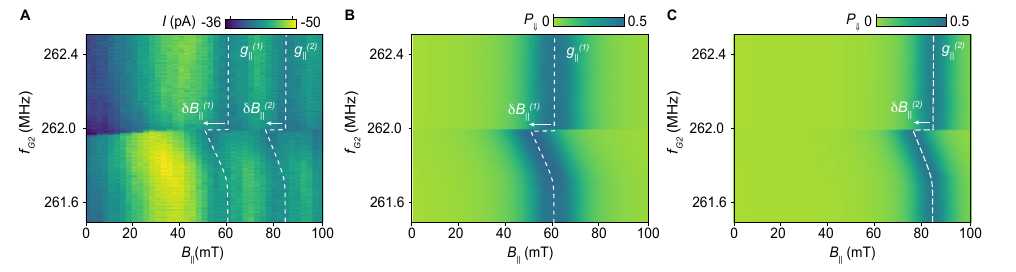} 
\caption[]{
\textbf{Resonant coupling between a single spin and mechanical motion}. 
(\textbf{A})~Measured resonant signal $I$ as a function of the applied parallel field $\Bpara$ and a single drive frequency $f_{G2}$ at power $P_{G2} = -33$~dBm. The white dashed lines are a guide to the eye for the EDSR resonances. For values of $f_{G2}$ far from $\fm$, these resonances corresponds to $g$-factors of $g_{\parallel}^{(1)}$ and $g_{\parallel}^{(2)}$.
At $f_{G2}$ slightly below $\fm = 261.9$~MHz, the EDSR resonances shift significantly towards lower values of $\Bpara$. 
White arrows indicate these shifts, $\delta \Bpara^{(1)}$ and $\delta \Bpara^{(2)}$. 
Here $I$ is detected by chopping the drive signal applied via $G2$ at 83.1~Hz and locking into the chopper signal~\cite{Laird2007}. 
(\textbf{B})-(\textbf{C})
Calculated transition probability $P_{\Downarrow}$ from the blocked to the unblocked state, time-averaged until the steady state is reached. The transition probability is obtained by solving the dynamics of Hamiltonian (2) up to second order in $\amp$ plus the EDSR driving tone, and including decoherence effects due to the environment (see Methods for details). 
Our simulation for the $g_{\parallel}^{(1)}$ and $g_{\parallel}^{(2)}$ resonances reproduce the EDSR resonances observed in panel \textbf{A}.}
\label{fig_3}
\end{figure*}
%%%%%%%%%%%%%%%%%%%%%

In order to evidence the spin-mechanical coupling, we monitor the EDSR resonance while actuating the nanotube's motion (Fig.~\ref{fig_2}{\bfseries A}, {\bfseries B}). Specifically, we fixed $\fspin$ at 5.1 GHz, thus driving the EDSR resonance corresponding to $g_{\x}^{(2)}$ in the vicinity of $\Bx^\EDSR=200$~mT, and swept $\fosci$ in a range of frequencies around $\fm$. For large enough values of $\Posci$, we observe a broadening $\Delta\Bx$ of the EDSR resonance at $\fosci\sim \fm$. To quantify this broadening we identify the values of $\Bx$ for which the current drops from the maximum EDSR current value by a factor of $1/\sqrt{2}$. 
The dependence of the broadening $\Delta\Bx$ on the power $\Posci$ is discussed further in the Methods.
We observe that the broadening $\Delta \Bx$ starts at slightly lower frequencies and hits the maximum at $\fm$. In contrast, for frequencies slightly larger than $\fm$, no broadening is observed. This asymmetry becomes more evident as $\Posci$ increases, indicating the presence of non-linearities in the mechanical motion. For instance, mechanical resonance hardening, i.e. an increase of $\fm$, and the occurrence of current switches that might be revealing mechanical bistabilities, are further evidence that mechanical non-linearities are present (see Methods). It is important to note that for Fig.~\ref{fig_2}\textbf{A}, \textbf{B}, $\fosci$ is swept from high to low frequencies. 

We model the impact of the interaction between the mechanics and the spin-valley qubit as a modulation of the qubit frequency $\fEDSR$.
The qubit is subjected to the applied magnetic field $\Bx$ with components parallel ($\Bpara$) and perpendicular ($\Bperp$) to the CNT's axis at zero displacement. 
The presence of spin-orbit coupling may induce an anisotropic gyromagnetic tensor~\cite{Payli2012}, $\mathbf{g}^{(j)}$, with components $\gpara^{(j)}$ and $\gperp^{(j)}$, where the superscript j indicates the different spin-valley resonances observed in figure~\ref{fig_1}\textbf{E} and their associated resonances.
For the CNT at rest, the qubit Hamiltonian for a given pair of spin-valley states is
\begin{equation} \label{eq:Hrest}
    H_{\rm rest}  = \frac{1}{2}\mu_{\mathrm{B}}\sqrt{\Bpara^2\gpara^2 + \Bperp^2\gperp^2} \, \sigth,
\end{equation}
where $\sigth$ is the Pauli operator in the energy quantisation axis of the qubit.
When the mechanical motion is driven, the CNT displaces with an amplitude $\displ$ and $\Bpara$ and $\Bperp$ in Eq.~\eqref{eq:Hrest} can be modified as follows: $\Bpara \to \left(\Bpara \, \ell - \Bperp \, \displ\right)/\sqrt{\displ^2 + \ell^2}$, and $\Bperp \to \left(\Bperp\,\ell + \Bpara\,\displ \right)/\sqrt{\displ^2 + \ell^2}$,
where $\ell$ is the distance between the qubit and the closest lead.
For small oscillation amplitudes $\displ$, we then expand to second order and obtain the qubit Hamiltonian 
\begin{equation} \label{eq:Hexpansion}
    H = H_{\rm rest}+ \lambda\ind{1} \, \displ \, \sigth + \lambda\ind{2} \, \displ^2 \, \sigth,
\end{equation}
where $\lambda\ind{1}$ and $\lambda\ind{2}$ are corrections to the qubit frequency that depend on $\gpara$, $\gperp$, and the field orientation (see Methods).

Using Eq.~\eqref{eq:Hexpansion} we can calculate the steady-state qubit transition probability $\Pflip$ to the unblocked state that is proportional to the current flowing through the CNT, $I$.
Given the signatures of mechanical non-linearities observed in Figs.~\ref{fig_2}\textbf{A} and \textbf{B}, we consider a Duffing model for the mechanics, see Methods.
Fig.~\ref{fig_2}\textbf{C} shows $\Pflip$ estimated from the Hamiltonian corresponding to Eq.~\eqref{eq:Hexpansion} up to first order, with $\lambda\ind{1} = 433 $ neV/nm.
Including just the first order correction in $\displ$ the model captures the physics of the off-resonant experiment, see Fig.~\ref{fig_2}\textbf{B}. 
We thus conclude that the main effect of the off-resonant coupling between the spin-valley qubit and the mechanical motion is a broadening of the EDSR resonance. The broadening of $\fEDSR$ is governed by the product of the coupling strength and the mechanical amplitude $\lambda\ind{1} \, \displ/h$, where $h$ is the Planck constant.

So far we have demonstrated spin-mechanical coupling by measuring the broadening $\Delta \Bx^\EDSR$ of the EDSR peak at $\fm$, see Fig.~\ref{fig_2}. This was achieved in an off-resonant regime, $\fEDSR>\fm$.
The model developed to account for spin-mechanical coupling, see Eq.~\eqref{eq:Hexpansion}, considers a modulation of the Zeeman Hamiltonian that depends on the magnetic field orientation and the g-factor anisotropy. 
To further test the nature of this coupling, we now align the magnetic field with the CNT axis. Our model in Eq.~\eqref{eq:Hexpansion} predicts that the first order correction $\propto \lambda\ind{1}$ is switched off and the Zeeman energy modulation is now dominated by the term quadratic in $\displ$ (see Methods). 
Moreover, for this magnetic field orientation, we are able to access the resonant spin-mechanical coupling regime, i.e. $\fEDSR\sim\fm$. 
This is because in this orientation, the g-factor's value is reduced, thus shifting the EDSR resonance condition for $\fm$ to higher fields where Pauli spin blockade is not lifted.
At the resonance condition, we can drive \emph{both} the EDSR resonances and the mechanical motion with a single tone applied to G2, $f_{G2}$ ($\fosci=\fspin=f_{G2}$). 

In Fig.~\ref{fig_3}\textbf{A} we display the EDSR resonances corresponding to g-factors $g^{(1)}_{\para}=0.31\pm0.03$ and $g^{(2)}_{\para}=0.22\pm0.02$
(these are the g-tensor components aligned with the CNT axis, while those in Fig.~\ref{fig_1}\textbf{E} are the $\x$-aligned components). 
For $f_{G2}\sim\fm$, we observe a shift of the EDSR resonances, $\delta \Bpara^{(j)} = \Bpara^{(j)}(\fm+\epsilon)-\Bpara^{j}(\fm-\epsilon)$ for small $\epsilon >0$ with $j=1,2$.
These shifts show a strong asymmetry:
at frequencies $f_{G2}$ smaller than $\fm$, the EDSR resonance is shifted to lower field strengths, reaching the maximum shift at $\fm$. In contrast, for frequencies only slightly larger than $\fm$, no shift is observed. (Note that for Fig.~\ref{fig_3}\textbf{A}, $f_{G2}$ is swept from high to low frequencies.)

To model the observed frequency shift, we now require in Eq.~\eqref{eq:Hexpansion} the second-order correction that arises from the spin-mechanical coupling. For each EDSR resonance, this is given by
\begin{equation} \label{eq:lambda2para}
    \lambda\ind{2} = E_\mathrm{rest}
    \frac{\gperp^2 - \gpara^2}{4\ell^2\gpara^2},
\end{equation}
where $E_\mathrm{rest}$ is the energy gap of $H_{\rm rest}$, see Eq. (\ref{eq:Hrest}) and Methods for details.
The time average effect of this coupling is to produce an effective qubit frequency given by
\begin{equation} \label{eq:1tonefshift}
    f_{\mathrm{eff}} = \frac{\mu_\mathrm{B}}{h}\gpara\Bpara + \frac{1}{2h} \lambda\ind{2} \, A^2(f_\mathrm{G2}),
\end{equation}
where $A(f_\mathrm{G2})$ is the frequency response of the mechanical oscillator considering the Duffing non-linearity.
This expression allows us to extract the value $\lambda\ind{2} \, A_\mathrm{max}^2$, with $A_\mathrm{max}$ the maximum displacement, from the fit of the EDSR resonances from Fig.~\ref{fig_3}\textbf{A} (white dashed line).
Fig.~\ref{fig_3}\textbf{B} shows the simulation of the transition probability using the parameters obtained from the fit of Eq.~\eqref{eq:1tonefshift}. 
We find that the experimental measurements, Fig.~\ref{fig_3}\textbf{A}, are extremely well matched with the theoretical model, Fig.~\ref{fig_3}\textbf{B} and \textbf{C}. 
This agreement unequivocally proves that the observed frequency shift in the resonance is due to spin-mechanical coupling. 

The fit of $\lambda\ind{2}$ together with  Eq.~\eqref{eq:lambda2para} allows us to make a quantitative prediction of the $\gperp$ coefficient. 
Here one needs to include an estimate of the qubit's position $\ell$ on the CNT, which is ca. 900~nm long. 
For a reasonable range of $\ell$, i.e. $\ell$=50 -- 250 nm, we find $\gperp$=4 -- 24.
Note, that the coupling corrections $\lambda\ind{1}$ and $\lambda\ind{2}$ are field-orientation dependent. However, having found the two g-tensor values, $\gpara$ and $\gperp$, we can now determine these corrections for \emph{arbitrary} field orientations. 
In particular, for the magnetic field applied along the $\x$ direction, we find $\lambda\ind{1} = $ 380 -- 455 neV/nm. This result is consistent with the value of 433 neV/nm used in the simulation of Fig.~\ref{fig_2}\textbf{C}.

\medskip 

%Conclusion
Here we have demonstrated spin-orbit-mediated coupling of a spin-valley qubit to high-frequency motion using a suspended carbon nanotube device.
As seen in Figs.~\ref{fig_2} and \ref{fig_3},  coupling is possible both when the qubit and mechanics are off-resonant, as well as when they resonate at the same frequency. 
The excellent theory-experiment match allows us to fully map out the coupling mechanism at play.
By probing this coupling for two different orientations of the applied magnetic field, we uncover that the coupling strength is strongly affected by the g-factor anisotropy. 
We also give a first experimental estimate of the strength $\lambda\ind{1}$ of the spin-mechanical coupling.

The spin-mechanical coupling we demonstrate unlocks a realm of experiments combining single spin qubits with a linear or non-linear, classical or quantum resonator. 
A single spin coupled to a quantum resonator acting as a battery allows, for example, the exploration of quantum batteries~\cite{Binder2015,Llobet2015,Culhane2022}, quantum Maxwell demons~\cite{Cottet2017} and thermodynamic engines~\cite{Josefsson2018}.
For highly coherent spin states~\cite{Cubaynes2019}, spin-mechanical coupling would allow for work extraction from quantum coherence~\cite{Kammerlander2016} and the realisation of macroscopic states of motion~\cite{Yiwen2023}. It could also enable ground-state cooling of a macroscopic resonator by spin-polarized currents~\cite{Stadler2014}.
The coupling of high-frequency mechanics and spin offers novel approaches for high-efficiency microwave-to-optical conversion~\cite{Forsch2020} and long-range coupling of spin qubits~\cite{Lukin2010,Lukin2021}. Mechanical resonators are smaller in size compared to their superconducting counterparts and their properties are not degraded in the presence of strong magnetic fields, enabling the realisation of large networks consisting of interconnected spin qubit registers.

%%%%%%%%%%%%%%%%%%%%%%%%%%%%%%%%%%%%%%%%%%%%%%%%%%%%%%%%%%%%%%%%%%%%%%%%%%%%%%%%%%%%%%%%%%%%%%%%%%%%%%%%%%%%%
\section*{acknowledgments}
We thank D. Zumbh\"{u}l, E. Laird and G. J. Milburn for stimulating discussions on the subject of this research. This research was supported by grant number FQXi-IAF19-01 from the Foundational Questions Institute Fund, a donor-advised fund of the Silicon Valley Community Foundation. NA acknowledges the support from the Royal Society (URF-R1-191150), EPSRC Platform Grant (grant number EP/R029229/1) and the European Research Council (ERC) under the European Union’s Horizon 2020 research and innovation programme (grant agreement number 948932). AP Acknowledges support from the HUN-REN Hungarian Research Network, by the Ministry of Culture and Innovation and the National Research, Development and Innovation Office (NKFIH) within the Quantum Information National Laboratory of Hungary (Grant No. 2022-2.1.1-NL-2022-00004), and by NKFIH via the OTKA Grant No. 132146. JM acknowledges funding from the Swedish Vetenskapsr\r{a}det, Swedish VR (Project No. 2018-05061). JA acknowledges support from EPSRC (EP/R045577/1), and thanks the Royal Society for support.
%%%%%%%%%%%%%%%%%%%%%%%%%%%%%%%%%%%%%%%%%%%%%%%%%%%%%%%%%%%%%%%%%%%%%%%%%%%%%%%%%%%%%%%%%%%%%%%%%%%%%%%%%%%%%

\section*{Methods}

\setcounter{figure}{0}
\renewcommand{\figurename}{Supplementary Fig.}

\subsection*{Device fabrication}
The carbon nanotube (CNT) used in this experiment has been grown by chemical vapor deposition and then transferred via flip-chip stamping on a silicon chip. For the CNT synthesis, we deposit a FeCl3 catalyst mixed with PMMA on a quartz substrate with etched pillars. 
The CNT are then grown in a furnace with a CH4/H2 atmosphere diluted to 20$\%$ concentration in Argon at 950 $^{\circ}$C. After the growth, the CNTs are transferred to the device chip by stamping using an optical mask aligner.
The device layout consisting of two 110 nm tall Cr/Au pillars and five 18 nm tall Cr/Au gate electrodes was realised on the silicon substrate using standard electron beam lithography followed by thermal evaporation of Cr/Au gates. All five gates underneath the CNT are connected to bias tees that allow the application of DC and high bandwidth AC voltages. 
The length and the radius of the CNT used for this experiment have not been measured directly, however these parameters have been estimated from previous experiments and found to be, respectively, $936 \pm 10$ nm and $3.9 \pm 0.2$ nm, see Ref.~\cite{Vigneau2022}.
The device was measured inside a Triton 200 cryofree dilution refrigerator with a base temperature of 45 mK, and equipped with a vector magnet that could generate 6 T along the cryostat main axis, defined as Y in the laboratory frame of coordinates, and 1 T along the remaining axis X and Z.

%%%%%%%%%%%%%%%%%%%%%
\begin{figure*}[t]
\includegraphics[trim=0.2cm 0.0cm 1.9cm 0.0cm,clip,width=1\textwidth]{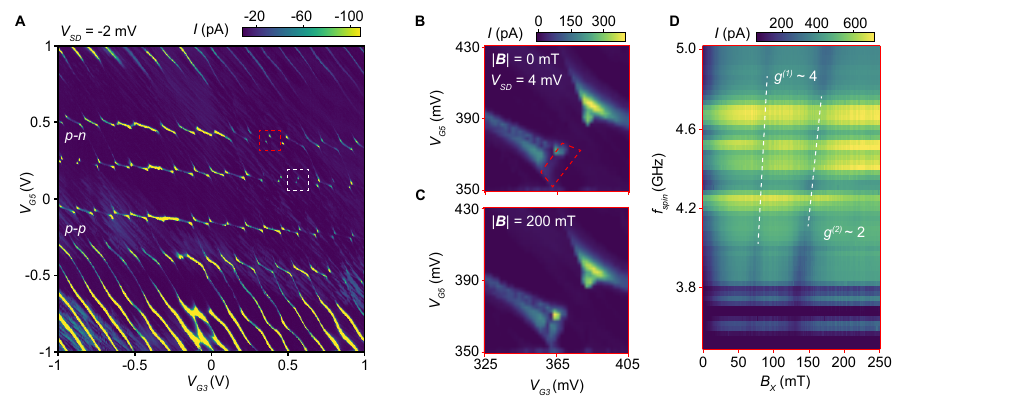}
\caption[]{\textbf{Pauli spin blockade and EDSR spectrum for a different charge transition.}
(\textbf{A}) Charge stability diagram of the double quantum dot measured as a function of $V_{G3}$ and $V_{G5}$ with $V_{SD} = -2$~mV. As a function of $V_{G5}$ we can tune the system into a double quantum dot. 
The white dash box indicates the pair of bias triangles used in the main text (see Fig.~$1$\textbf{B}-\textbf{C}), which shows Pauli blockade at negative source-drain bias. Similarly, in panels \textbf{B} and (\textbf{C} we show the current measured with positive bias $V_{SD} = +4$~mV, probed in the vicinity of the bias triangles at the (N-1, 2) $\to$ (N, 1) transition (see red dashed box), at different magnetic fields magnitudes $|\bm{B}|=0$~mT and $|\bm{B}|=200$~mT respectively.
At zero magnetic field the current at the base of the lower left bias triangle is suppressed by Pauli blockade, see the red dashed box in panel \textbf{B}. (\textbf{D}) EDSR spectrum as a function of the drive frequency $\fspin$ and applied field $\Bx$. Like the data presented in the main text, we identify two EDSR resonances as regions of suppressed current with $g$-factors respectively $g^{(1)}\approx 4$ and $g^{(2)}\approx 2$ (see white dashed lines).
}
\label{S1}
\end{figure*}
%%%%%%%%%%%%%%%%%%%%%

%%%%%%%%%%%%%%%%%%%%%
\begin{figure*}[t]
\centering
\includegraphics[trim=3.4cm 0.0cm 3.4cm 0.0cm,clip,width=0.7\textwidth]{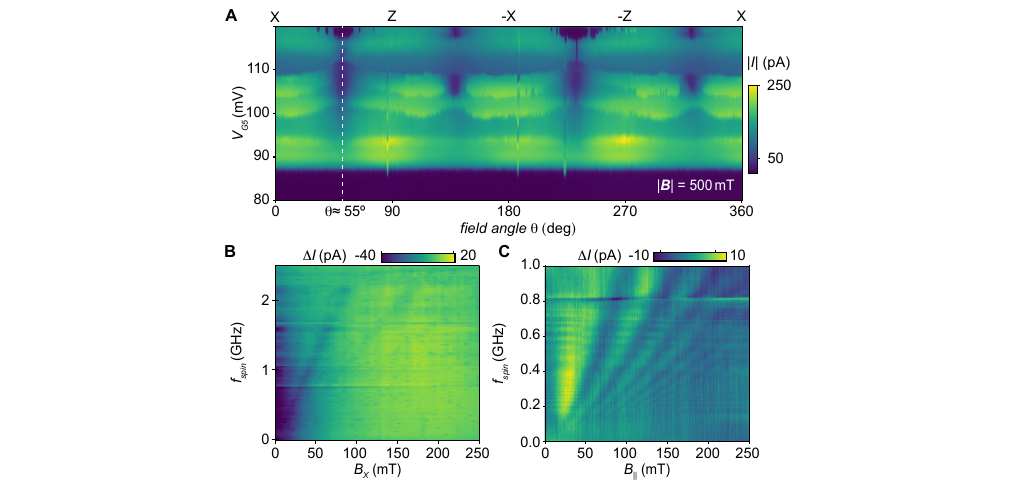}
\caption[]{\textbf{Magneto transport angular dependence and EDSR at different field orientations.} (\textbf{A}) Current measured as a function of $V_{G5}$ and the angle $\theta$ which determine the magnetic field orientation in the $XZ$ plane in our frame of coordinates. The magnitude of the magnetic field was kept constant at 500 mT. Supported by geometrical considerations, we identify these regions of suppressed current with the directions for which the field aligns along the main carbon nanotube axis. The angle $\theta$ = 55$^\circ$, highlighted by a white dashed line marks the orientation that we called $\Bpara$ in the rest of the manuscript. (\textbf{B}) EDSR resonances measured with the field aligned along the $X$-axis of our reference frame (adapted from Fig.~1\textbf{E} of the main paper). (\textbf{C}) EDSR resonances measured at the field orientation $\Bpara$.
}
\label{S2}
\end{figure*}
%%%%%%%%%%%%%%%%%%%%%

\subsection*{Forward bias EDSR and Qubit location}
To confirm the EDSR signal, we measured it at a different charge transition with opposite source drain bias.
In Extended Data Fig.~\ref{S1}{\bfseries A} we present a stability diagram. As a function of $V_{G5}$, starting from the lower corner, we observe a transition to a double dot regime suggesting the presence of a bandgap for the right dot (highlighted by a white shade)~\cite{Laird2015}. We therefore identify two transport-type regions p-p and p-n, which allow us to assign the p-n charge occupation (N, 1)-(N-1, 0) to the pair of bias triangles shown in Fig.~1{\bfseries B}-{\bfseries C} of the main text and here highlighted by the dashed white box. As shown in the main text, this transition exhibits Pauli-blockade when transport is measured using a negative source-drain bias. On the other hand, the pair of triangles highlighted by the dashed red box marks the transition (N+1, 2) $\to$ (N, 1). 
Extended Data Fig.~\ref{S1}{\bfseries B-C} shows Pauli blockade at this particular pair of bias triangles measured with a source-drain bias of $V_{SD} = +4$~mV. Similar to Fig.~1{\bfseries B} of the main text, at $\Bx=0$~mT, the current at the base of the lower left bias triangle is suppressed by Pauli blockade, see the red dash box in Extended Data Fig.~\ref{S1}{\bfseries B}. Measuring the current within this region in the presence of a continuous wave excitation $\fspin$, reveals the EDSR spectrum presented in Extended Data Fig.~\ref{S1}{\bfseries D}.
Consistent with the data presented in the main text, even in forward bias the two observed EDSR resonances are evident as regions of suppressed currents with estimated $g$-factors of 2 and 4 respectively. We find that this observation, considering that no EDSR was detected outside either of the Pauli-blockaded regions, confirms that we can define spin qubits in our CNT device.

\subsection*{Magneto transport spectroscopy and EDSR as a function of field orientation}
Extended Data Fig.~\ref{S2}{\bfseries A} shows magneto transport spectroscopy measured as a function of gate voltage $V_{G5}$ and field orientation in the X-Z plane of our reference frame (see Fig.~1{\bfseries A} in the main text.) Throughout this measurement, we fixed the magnetic field magnitude to 500 mT, and set $V_{G3}$ in order to probe the region of current rectified by Pauli blockade as a function of $V_{G5}$. As a function of the field angle $\theta$, we observe two regions of suppressed current with 180$^{\circ}$ periodicity at angles $\theta$ = 55$^{\circ}$ and 235$^{\circ}$, and angles 145$^{\circ}$ and 235$^{\circ}$. 
We attribute the regions of strongest (weakest) current suppression to the direction for which the field aligns parallel or anti-parallel with (perpendicular to) the main axis of the carbon nanotube. Extended Data Fig.~\ref{S2}{\bfseries B} shows EDSR resonances measured with the field oriented along the $X$ direction of the reference frame. Extended Data Fig.~\ref{S2}{\bfseries C} shows EDSR resonances measured with the field oriented along the direction of strongest current suppression, which we labeled $\Bpara$ (see the white dashed line in panel {\bfseries A} at $\theta$ = 55$^{\circ}$). 

% \newpage 

\subsection*{Power dependence of the spin mechanical coupling}
\begin{figure*}[t]
\includegraphics[trim=0.2cm 0.1cm 0.2cm 0.0cm,clip,width=1\textwidth]{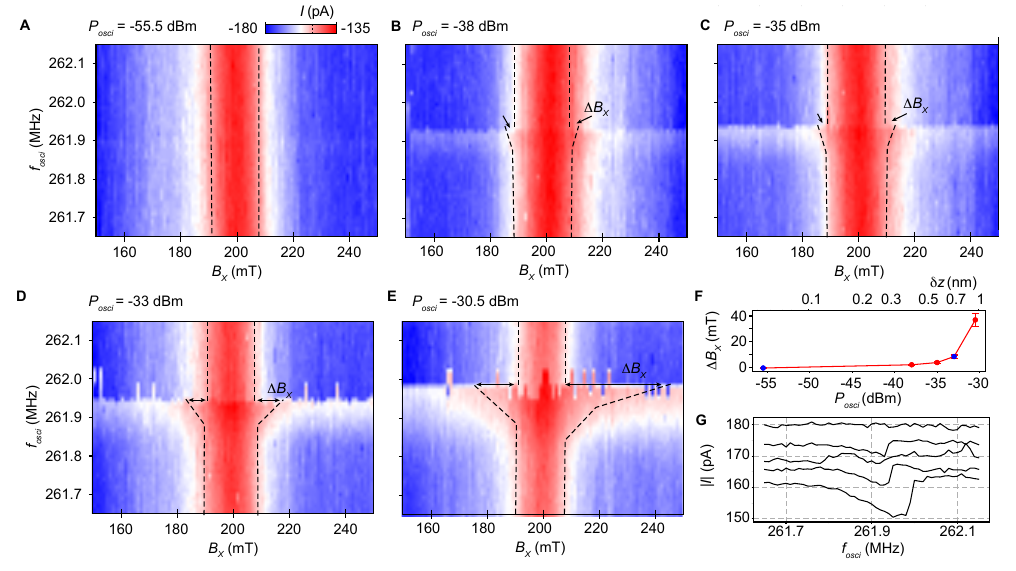}
\caption[]{
%Continuous wave EDSR resonance at a field $\Bx\approx 200$~mT with spin drive $\fspin \sim 5.1$~GHz and power $\Pspin=-34.5$~dBm. The CNT motion is excited (at G2) at frequencies $\fosci$ in the range [261.7, 262.1]~MHz. 
\textbf{Power dependence of the spin mechanical coupling.} (\textbf{A})-(\textbf{E}) Non-resonant spin mechanical coupling for increasing mechanical drive power $\Posci$, from -55.5 dBm to -30.5 dBm. Panel \textbf{A} and \textbf{D} are the same as Fig.~2 \textbf{A}-\textbf{B} in the main text. With increasing values of $\Posci$, the EDSR resonance broadens ($\Delta \Bx$ indicated by arrows) while approaching the mechanical resonance frequency $\fm$ (see panel \textbf{F} top). Seemingly, with increasing driving power the mechanical resonance shifts to higher frequencies due to the non-linearity of the mechanical oscillator. (\textbf{F}) Broadening of the EDSR resonance peak at $\fosci \sim \fm$ as a function of the mechanics driving power $\Posci$. 
%To obtain $\Bx$ we estimate when the current $I$ has dropped from the maximum EDSR current value by a factor of $1/\sqrt{2}$ (white dashed lines). 
The blue dots indicate the estimated broadening of the data presented in the main text. On the top axis, we report the overall nanotube displacement $\delta$z estimated from $\Posci$ using a standard electrostatic model with the CNT mass, physical dimensions, and capacitance to the electrostatic gates are obtained from previous experiments~\cite{Vigneau2022}. (\textbf{G}) Linecuts extracted from \textbf{A} to \textbf{E} going from top to bottom (traces are offset for clarity), at $\Bx$ = 185 mT, showing the current as a function of $\fosci$, for different excitation powers.}
\label{S3}
\end{figure*}

In Extended Data Fig.~\ref{S3}, we show additional data displaying the off-resonant spin mechanical coupling as a function of the mechanical drive power $\Posci$. From panels {\bfseries A} to {\bfseries E}, $\Posci$ is stepped from $-55.5$ dBm to $-30.5$ dBm, with panels {\bfseries A} and {\bfseries D} corresponding respectively to the data presented in Fig.~2{\bfseries A} and 2{\bfseries B} of the main text. With increasing $\Posci$, we observe a broadening ($\Delta\Bx$) of the EDSR resonance when $\fosci \sim \fm$. The broadening is obtained by comparing the width of the EDSR at $\fosci = 261.65$ MHz, with its width at $\fosci = \fm$. Here, the width is defined as the range of field values where the current $I$ is within a factor $1/\sqrt{2}$ larger than the minimum EDSR current (see white dashed lines). In panel {\bfseries F} top we plot the estimated ($\Delta\Bx$) divided by 2, as a function of the drive power $\Posci$ and the corresponding carbon nanotube displacement $\delta$z. As discussed in the main text, the broadening shows a non-linear dependence on the driving power, an effect of the nonlinear nature of the oscillator.
In Extended Data Fig.~\ref{S3}{\bfseries G} we present further evidence for the nonlinearity of the mechanical oscillator.   
The current traces as a function of $\fosci$ are linecuts extracted from Extended Data Fig.~\ref{S3}{\bfseries A}-{\bfseries E} and highlight the hardening of the mechanical frequency $\fm$ as a function of $\Posci$.  
Previous studies on CNT resonators have found these signatures to be consistent with a Duffing oscillator~\cite{Steele2009,Gomez2012}.
Recently, Duffing oscillators in CNT resonators have been proposed as a platform to realise nanomechanical qubits~\cite{Pistolesi2021}.

%%%%%%%%%%%%%%%%%%%%%
%%%%%%%%%%%%%%%%%%%%%

%%%%%%%%%%%%%%%%%%%%%
\begin{figure*}
\includegraphics[trim=0.4cm 0.1cm 0.4cm 0.2cm,clip,width=1\textwidth]{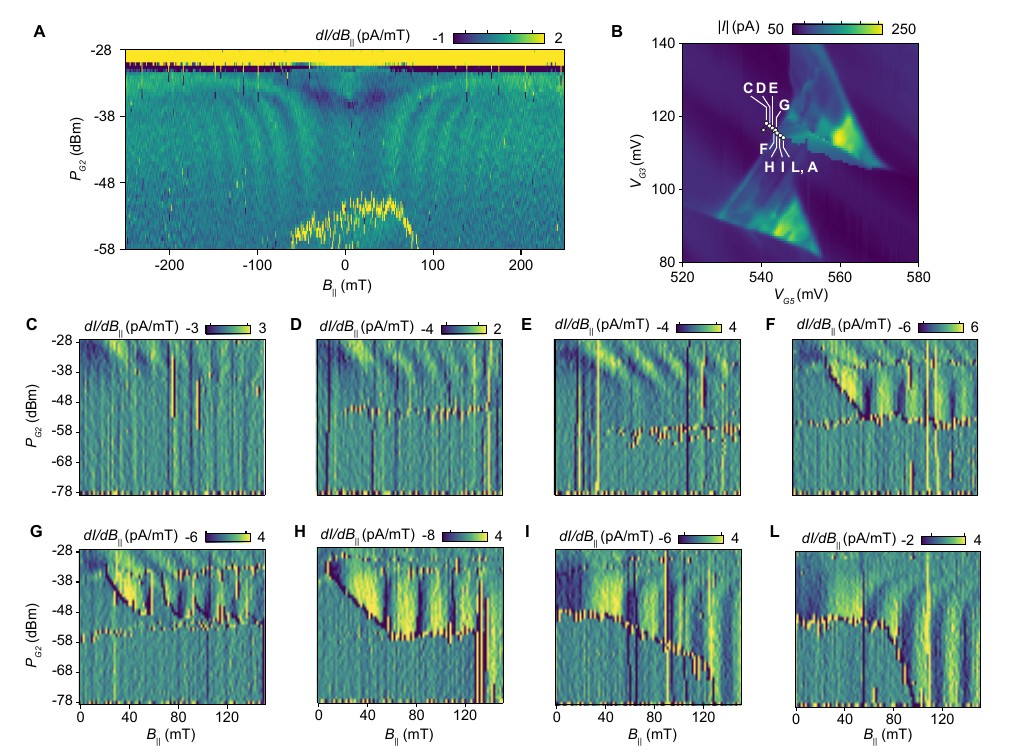}
\caption[]{
\textbf{Detuning dependence of the spin-mechanical coupling.} (\textbf{A}) here we show $dI/d\Bpara$ revealing EDSR resonances as a function of applied parallel field $\Bpara$ and drive power $P_{G2}$ at a fixed drive frequency $f_{G2} = 261.9$ MHz. For increasing values of $P_{G2}$ the EDSR resonances shift significantly towards higher values of $\Bpara$. (\textbf{B}) Charge stability diagram of the double dot with markers indicating the detunings at which panels \textbf{A} and \textbf{C}-\textbf{L} have been measured. The star indicates Fig.~3\textbf{A} in the main text. Panels (\textbf{C})-(\textbf{L}) show EDSR resonances measured as a function of $P_{G2}$ and $\Bpara$ at $f_{G2} = 261.9$ MHz. Each panel is measured at a different detuning point highlighted by circles in panel \textbf{B}. Like in panel \textbf{A}, we observe a shift in the EDSR resonances as a function of $P_{G2}$. Similar to Fig.~3\textbf{A} in the main text, in panels \textbf{C} to \textbf{G} the observed shift occurs towards lower values of $\Bpara$. In panel \textbf{H} we do not observe any appreciable shift, while similar to panel \textbf{A}, panels \textbf{I} and \textbf{L} show the resonances shifting towards higher values of the applied field.  
}
\label{S4}
\end{figure*}
%%%%%%%%%%%%%%%%%%%%%

\subsection*{Detuning dependence of the spin mechanical coupling}
In Extended Data Fig.~\ref{S4}{\bfseries A} we show EDSR resonances measured as a function of drive power $P_{G2}$ and applied field $\Bpara$ with a single drive frequency such that $ \fosci = \fspin = f_{G2} = \fm = 261.9$ MHz. This allows us to drive the spin while simultaneously driving the mechanics and investigating the coupling as a function of the driving powers. The fixed frequency driving reveals a fan of EDSR resonances. As a function of $P_{G2}$ we observe a shift in the EDSR resonances towards higher values of the applied field $\Bpara$. We note that this is in contrast with the data presented in Fig.~3{\bfseries A} of the main text, where the observed EDSR resonances shift towards lower values of the applied field as a function of the drive frequency.

To investigate this further, we take similar measurements at different detuning values. Extended Data Fig.~\ref{S4}{\bfseries B} shows the charge stability with circles indicating the detuning corresponding to Extended Data Fig.~\ref{S4}{\bfseries A} and Extended Data Fig.~\ref{S4}{\bfseries C} to {\bfseries L}. Note that the detuning of Fig. 3{\bfseries A} in the main text is highlighted with a star. 
As a function of detuning, we observe a distinct change in the sign of the EDSR resonant shift, with panels {\bfseries C}-{\bfseries G} exhibiting a shift towards lower values of the applied eternal filed, panel {\bfseries H} showing no appreciable shift, and panels {\bfseries I}, {\bfseries L}, and {\bfseries A} showing a marked shift toward higher values of magnetic field. 
This change in the sign of the spin-resonance shift indicates that we could in principle tune the coupling between the spin and the mechanical motion. We attribute this tunability to the modulation of the g-tensor components with gate voltages $V_{G3}$ and $V_{G5}$.

The model employed to explain the origin of the spin--mechanical coupling suggests that in the resonant case, the coupling parameter $\lambda\ind{2}$ depends on the difference of the squares of the g-factors components, $\gperp$ and $\gpara$.
This can in principle lead to variable coupling strength owing to a modulation of the g-factor components with detuning. $g$-tensor modulation with gate voltages and applied magnetic field has been observed in a variety of materials including SiGe quantum dots~\cite{Ares2013}, Ge/SiGe heterostructures~\cite{Jirovec2022}, and CMOS silicon on insulator quantum dots devices~\cite{Michal2023,Crippa2018}.

\subsection*{Theoretical model}
The Hamiltonian of the spin-valley qubit in the presence of a magnetic field $\mathbf{B}$ is in general given by
\begin{equation}
    H_0 = \frac{1}{2}\muB\mathbf{B}\cdot\mathbf{g}\cdot\bm{\sigma},
\end{equation}
where $\mathbf{g}$ is the anisotropic g-tensor~\cite{Flensberg2010,Sen2023}.
The off-and-on resonance experiments described in the main text were performed with two different field orientations on a plane containing the CNT.
As such, without loss of generality, we define a parallel ($\Bpara$) and perpendicular ($\Bperp$) component of the magnetic field to the CNT at rest, so that the third component of $\mathbf{B}$ is zero for all measurements.
The Hamiltonian for the CNT at rest is then given by
\begin{align}
    H\rest &= \frac{1}{2}\muB(\Bpara\gpara\sigma_{\para} + \Bperp\gperp\sigma_\perp)
    \\ \nonumber
    &= \frac{1}{2}\muB\sqrt{\Bpara^2\gpara^2 + \Bperp^2\gperp^2}\,\sigma_3 = \frac{1}{2}E\rest\sigma_3,
\end{align}
where $\sigma_3$ is the Pauli operator in the direction of the energy quantisation axis of the qubit.

If the CNT oscillates within this same plane defined by magnetic field orientations $\Bpara$ and $\Bperp$, then under a displacement $\chi$ of the CNT, the parallel and perpendicular field components, with respect to the CNT main axis, rotate by an angle $\tan\theta = \ell/\chi$, where $\ell$ is the distance of the quantum dot to the closest lead (see Fig.~$1${\bfseries A} of the main text.).
The parallel and perpendicular components of the field change according to
\begin{equation}
    \Bpara \longrightarrow \frac{\Bpara\,\ell - \Bperp\,\chi}{\sqrt{\displ^2 + \ell^2}},
    \qquad
    \Bperp \longrightarrow \frac{\Bperp\,\ell + \Bpara\,\chi}{\sqrt{\displ^2 + \ell^2}}.
\end{equation}
The Hamiltonian in the new quantisation axis is then
\begin{equation}
    H_0 = \frac{1}{2}\muB\sqrt{\frac{(\Bpara\,\ell - \Bperp\,\chi)^2\gpara^2}{\chi^2 + \ell^2} + \frac{(\Bperp\,\ell + \Bpara\,\chi)^2\gperp^2}{\chi^2 + \ell^2}}\,\sigma_3.
\end{equation}
Expanding to the second order in $\chi$ we then have
\begin{equation} \label{eq:sm:Hexpansion}
    H_0 = \frac{1}{2}E_{\rest}\sigma_3 + \lambda_1\chi\sigma_3 + \lambda_2\chi^2\sigma_3,
\end{equation}
where
\begin{align} \label{eq:sm:lambdas}
    \lambda_1 &= \muB^2\frac{\Bperp\Bpara\left(\gperp^2-\gpara^2\right)}{2\ell E\rest},
    \\
    \lambda_2 &= \muB^4\frac{\left(\Bpara^4\gpara^2 - \Bperp^4\gperp^2\right)\left(\gperp^2-\gpara^2\right)}{4\ell^2E\rest^3}.
\end{align}

We will here consider the mechanics as a classical system and hence $\chi$ will be a classical displacement satisfying the equations of motion of a Duffing oscillator~\cite{Steele2009,Gomez2012,Pistolesi2021}, that is
\begin{equation}
    \ddot{\chi} + \nu\dot{\chi} + \omega_m^2\chi + \beta\chi^3 = \frac{F_0}{m}\cos(\omega_dt),
\end{equation}
where $\nu$ is the damping coefficient of the mechanics, $\beta$ the Duffing coefficient, $F_0$ the driving force, and $m$ the mass of the CNT.
To deal with the Duffing equation analytically or numerically it is very convenient to rewrite the equations of motion in dimensionless quantities
\begin{align}
    \bar t &= t\omega_m, \quad
    \bar\omega = \frac{\omega}{\omega_m}, \quad
    \bar\chi = \frac{m\omega_m^2}{F_0}\chi, \quad
    \\ \nonumber
    \bar\nu &= \frac{\nu}{\omega_m}, \quad
    \bar{\beta} = \frac{\beta F_0^3}{m^3 \omega_m^8}.
\end{align}
so that
\begin{equation}
    \ddot{\bar\chi} + \bar\nu\dot{\bar\chi} + \bar\chi + \bar\beta{\bar\chi}^3 = \cos(\bar\omega_d\bar{t}).
\end{equation}
In what follows we will focus on the case of very weak non-linearity, $\bar\beta \ll 1$. In this regime, the response of the oscillator is approximately harmonic, with an amplitude given by the frequency-dependent Duffing response. The values of $\bar\beta$ extracted from the experimental data and shown in the next section, are consistent with this assumption.

\medskip

The qubit is further driven by the EDSR pulse, which induces Rabi oscillations. Assuming the condition for the rotating wave approximation (RWA) we then have the effective qubit Hamiltonian
\begin{equation}
    H = H_0 + \Omega\sigma_1,
\end{equation}
where $\Omega$ is the Rabi frequency.

\medskip

Finally, the qubit in the experiment is subject to different sources of noise that induce strong decoherence. To account for this effect, we use a standard pure--dephasing master equation
\begin{equation} \label{eq:sm:me}
    \dot{\rho} = -\frac{i}{\hbar}[H,\rho] + \Gamma\left(\sigma_3\rho\sigma_3 - \rho\right),
\end{equation}
where $\Gamma$ determines the decoherence time.
For the case where decoherence is dominated by pure dephasing~\cite{Pei2017}, we have that $\Gamma \sim 1/T_2^*$.

\subsubsection*{Off-resonant case}
In the off-resonant case, the magnetic field is applied in the direction $\x$, forming an angle $\theta \approx 55^\circ$ with respect to the direction parallel to the CNT at rest. To model these measurements, we use Eq.~\eqref{eq:sm:me} with $H_0$ expanded to first order in $\chi$, see Eq.~\eqref{eq:sm:Hexpansion}.
The simulation of Fig.~2{\bfseries C} of the main text is done by time evolving the master equation \eqref{eq:sm:me} until steady-state is reached and time-averaging the unblocked state probability. For the simulation, the
parameters taken are
\begin{align}
    &\lambda_1^{(2)} = 433 \,\frac{\mathrm{neV}}{\mathrm{nm}},
    \quad
    \bar\beta = 2\times 10^{-12},
    \\ \nonumber
    &\Gamma = 392.9 \,\mathrm{MHz}, 
    \quad
    \Omega = 2\pi \times 7.6 \,\mathrm{MHz}.
\end{align}
The spin-mechanics coupling $\lambda_1^{(2)}$ is chosen to match the observed EDSR broadening observed at the mechanical resonance.
The value of $\bar\beta$ is set to match the observed asymmetry in the frequency response of the mechanical oscillator, while the values of $\Omega$ and $\Gamma$ are chosen to match the EDSR width.
It is worth noting that within the regime of strong decoherence, $\Gamma \gg \Omega$, particularly precise values $\Omega$ and $\Gamma$ are not too relevant. For the calculations corresponding to the simulation displayed in Fig.~2{\bfseries C} in the main text only the width of the resonance depends on their value.

\subsubsection*{Resonant, parallel field case}
In the resonant case, the magnetic field is oriented parallel to the CNT at rest. That is, $\Bperp = 0$, $\Bpara = B$. Substituting into Eq.\eqref{eq:sm:lambdas} we see that in this case $\lambda_1 = 0$, i.e. the linear contribution of the mechanics vanishes. Therefore, the lowest order contribution from the mechanics is the quadratic term, and the second coupling takes the simple form given in Eq.~(3) of the main text.

\medskip

As described in the main text, looking at the effective frequency shift produced by the mechanics in this case, we extract the analytical expression for the frequency shift
given in Eq.~(4) of the main text. We fix the values of the resonance $\omega_m/2\pi = 261.9$ MHz, and quality factor to $Q = 5000$ and fit the remaining parameters to the observed frequency shifts. We obtain
\begin{align}
    \bar\beta = 2\times 10^{-11},
    \qquad
    &\lambda_2^{(1)}\frac{\chi_\mathrm{max}^2}{\gpara^{(1)}\muB} = 9.6\,\mathrm{mT},
    \\ \nonumber
    &\lambda_2^{(2)}\frac{\chi_\mathrm{max}^2}{\gpara^{(2)}\muB} = 8.3\,\mathrm{mT},
\end{align}
where $\chi_\mathrm{max}$ is the maximum displacement of the CNT, the superscript indicates each of the two resonances observed in Fig.~2{\bfseries C} of the main text, and the values of $\gpara$ are reported in the main text.
It is worth noting that the value of the Duffing coefficient in this case is not exactly the same as the off-resonant one. This is not surprising as previous works have shown that the Duffing non-linearity in a CNT depends strongly on the applied gate voltages~\cite{Steele2009,Gomez2012}.

\medskip

From the parameters of the CNT and the mechanical driving power $\Posci$ we estimate a maximum displacement of $1.5\,\mathrm{nm}$, following the same procedure outlined in the supplementary material of Ref.~\cite{Vigneau2022} (with the same parameters). This gives us the second-order coupling constant
\begin{equation}
    \lambda_2^{(1)} = 76.5\,\frac{\mathrm{neV}}{\mathrm{nm}^2},
    \qquad
    \lambda_2^{(2)} = 54.3\,\frac{\mathrm{neV}}{\mathrm{nm}^2}.
\end{equation}
In terms of the zero point motion, the bare second order coupling $\kappa^{(k)}_2 = \lambda_2^{(k)}\chi_\mathrm{zpm}^2$ is
\begin{equation}
    \kappa^{(1)}_2 = 2\pi\times 8.5 \,\mathrm{Hz},
    \qquad
    \kappa^{(2)}_2 = 2\pi\times 6.0 \,\mathrm{Hz}.
\end{equation}

%%%%%%%%%%%%%%%%%%%%%%%%%%%%%%%%%%%%%%%%%%%%%%%%%%%%%%%%%%%%%%%%%%%%%%%%%%%%%%%%%%%%%%%%%%%%%%%%%%%%%%%%%%%%%

%%%%%%%%%%%%%%%%%%%%%%%

\end{document}